\documentclass[11pt]{article}
\usepackage[margin=1in]{geometry}
\usepackage{amsmath,amssymb}
\usepackage{graphicx}
\usepackage{booktabs}
\usepackage[hidelinks]{hyperref}
\usepackage{xcolor}
\usepackage{caption}
\newcommand{\Aa}{\text{\AA}}
\newcommand{\rc}{r_c}
\newcommand{\Rstar}{r^{*}}

\title{\bfseries An entropic bottleneck, dynamical gating, and outward redistribution\\
of roaming in a designed Chesnavich-type model}
\author{Stephen  Wiggins\\[2pt]
\small Hetao Institute of Mathematics and Interdisciplinary Sciences, Shenzhen, China\\
\small School of Mathematics, University of Bristol, Bristol BS8 1TW, United Kingdom}

\begin{document}
\maketitle

\begin{abstract}
\noindent
Roaming reactions are organized not by potential-energy saddles but by transition states that
are unstable invariant objects in phase space, periodic orbits in the two degrees of freedom
studied here. To ask what controls roaming, we modify the
Chesnavich model of a barrierless ion--molecule dissociation: its orientation-dependent angular
hindrance is replaced by a transverse-stiffness ridge whose angular frequency peaks at an
interior radius, and the classical dynamics are studied at a fixed energy just above the
dissociation threshold. Comparing two ensembles that differ only in this angular interaction
(same radial channel, energy, and inward initial conditions) isolates its effect. The ridge
gates entry into the inner well, cutting inner capture from $57\%$ to $15\%$ and returning most
of the incoming flux directly to reactants; it does not eliminate roaming but relocates it
outward, suppressing it inside the ridge and switching it on farther out. The model retains
analogues of the original model's three transition states (tight, free-rotor, and outer orbiting
orbits), which we locate as unstable periodic orbits. The tight orbit spans a dividing surface that coincides, within numerical
accuracy, with the variational minimum-flux surface, and it carries no barrier along the reaction
coordinate: a deep entropic
bottleneck placed at an interior radius by the stiffness maximum. Its entropic character is
shared with the original model. Strength-matched monotone controls show that the gating tracks
the hindrance strength at the bottleneck radius; what the interior maximum supplies is placement,
concentrating that strength where it gates most effectively. The trajectories it admits roam
nonstatistically, with nonexponential gap-time
distributions: the entropic bottleneck governs how much is captured, not the dynamics that
follow.
\end{abstract}

\section{Introduction}
\label{sec:intro}

\paragraph{Roaming and entropic transition states.}
Conventional transition-state theory places a locally nonrecrossing dividing surface at a
rank-one saddle of the potential energy surface. Two kinds of process fall outside that picture:
barrierless bond fissions, whose transition states must be found variationally because the
potential rises monotonically along the reaction coordinate, and roaming reactions, in which a
partially dissociated fragment executes large-amplitude motion in a flat long-range region and
then reacts or escapes without following the minimum-energy path. Since H-atom roaming was
identified in formaldehyde~\cite{townsend2004} it has been documented across many systems and is
now recognized as a general feature of gas-phase dynamics~\cite{suits2008,bowman2014,
bowmanshepler2011,suits2020,fernando2015}. For barrierless association and dissociation the
controlling loose transition state is located variationally~\cite{truhlar1996,georgievskii2005},
and its cost is predominantly entropic rather than tied to a saddle. A related but distinct line
of work concerns the ``entropic intermediate'' of nonstatistical organic dynamics, a free-energy
minimum on a flat region of the surface with long residence times and heavy
recrossing~\cite{tantillo2021}; we treat it here as an adjacent phenomenon rather than the same
physics. In phase space these transition states are not saddles but unstable invariant objects and the
dividing surfaces attached to them: periodic orbits in systems with two degrees of freedom, and
normally hyperbolic invariant manifolds in general~\cite{wsw2008,waalkenswiggins2010,wiggins2016}. The model studied
here has two degrees of freedom, so its transition states are periodic orbits, and roaming is
organized by a small set of them.

\paragraph{Entropy barriers.}
That a rate can be controlled by entropy rather than energy is long established: Zhou and
Zwanzig~\cite{zhouzwanzig1991} gave a solvable model in which diffusion through a channel of
varying width produces a purely entropic barrier. Most relevant here, Makarov~\cite{makarov2017}
showed that a localized maximum of a transverse frequency along a reaction coordinate generates
an entirely entropic free-energy barrier, one with no potential maximum along the coordinate,
because the constrained free energy
\begin{equation}
F(x)=V(x)+k_BT\ln\frac{\omega_\perp(x)}{\omega_0}
\label{eq:makarovF}
\end{equation}
acquires a maximum wherever the transverse mode stiffens, even where $V(x)$ is flat. Here $x$ is
the reaction coordinate, $V(x)$ the potential along it, $\omega_\perp(x)$ the transverse-mode
frequency at $x$, $\omega_0$ a reference frequency, and $k_BT$ the thermal energy. In the model
studied here the reaction coordinate is the radial separation $r$ and the transverse mode is the
angular coordinate $\theta$. Makarov further showed that such a barrier is dynamically distinct
from what controls crossing times: transition-path-time statistics can be insensitive to a
barrier that nonetheless dominates the free-energy profile. Both points recur below. The entropic
character of such a bottleneck is not itself new: in a barrierless system any variational
bottleneck is entropic, and the Chesnavich model we build on already has one. What we add is a
deliberately deep entropic bottleneck, placed at an interior radius by an interior maximum of the
transverse stiffness, embedded in a Hamiltonian model of roaming where its dynamical consequences
can be followed.

\paragraph{The Chesnavich model and the question.}
Chesnavich introduced a model Hamiltonian for ion--molecule dissociation to study competition
between multiple transition states~\cite{chesnavich1986,suchesnavich1982}; its planar,
zero-angular-momentum form is now a standard laboratory for roaming. Maugui\`ere, Collins, Ezra,
Farantos and Wiggins analyzed roaming in it in phase space, identifying the tight and orbiting
transition states with dividing surfaces attached to unstable periodic orbits and associating
roaming with free-rotor orbits born in saddle--center bifurcations~\cite{mauguiere2014jcp,
mauguiere2014cpl,mauguiere2015jpcl,mauguiere2016ozone,roamingARPC2017}; Kraj\v{n}\'ak and
Waalkens, and Kraj\v{n}\'ak and Wiggins, mapped these structures in detail and across parameter
space~\cite{KrajnakWaalkens18,KrajnakWiggins18,KrajnakWiggins24,EzraWiggins19,
KrajnakEzraWiggins19a,KrajnakEzraWiggins19b}. In that model the tight and orbiting transition
states are tied respectively to the orientation-locking angular hindrance and to the centrifugal
barrier. We ask a dynamical question. If the orientation-locking hindrance is replaced, on the
same barrierless radial channel, by a ridge in the transverse stiffness (an interior maximum of
the angular frequency, with no potential barrier along the channel), how is roaming changed? At
fixed energy, and for the same incoming ensemble, does the ridge promote roaming, suppress it, or
reorganize it? Our central finding is the last: the ridge gates entry into the inner well,
admitting only a small fraction of the incoming flux and returning most of it directly to
reactants, and it does not increase the roaming fraction but displaces it outward, suppressing it
inside the ridge and switching it on farther out.

\paragraph{Three transition states.}
The phase-space account of this behavior rests on three transition states, all unstable periodic
orbits located here at fixed energy: a tight orbit, a free-rotor orbit, and an outer
orbiting orbit. The tight orbit gates entry to the well; the dividing surface it spans coincides with the
variational minimum-flux surface (as verified in Sec.~\ref{sec:threeradii} by comparing the tight-orbit action with the
fixed-radius minimum flux) and carries no potential barrier along the
radial coordinate, so it is a purely entropic bottleneck, made deep and placed at an interior
radius by the transverse-stiffness maximum. The free-rotor orbit sorts direct from roaming
trajectories; the orbiting orbit gates escape to products. The tight orbit is the one the
ridge reshapes: its entropic character it shares with the Chesnavich tight transition state, but
the designed ridge makes it deep enough to control capture. The roaming it admits is
nonstatistical, with nonexponential gap-time distributions: the entropic bottleneck sets how much
is captured, not the fate of what follows.

\paragraph{Organization.}
Section~\ref{sec:chesnavich} fixes the classical Chesnavich model. Section~\ref{sec:design}
introduces the designed transverse-stiffness model. Section~\ref{sec:matched} presents the
central result: two trajectory ensembles compared under identical conditions (same radial
potential, energy, and incoming ensemble), differing only in the angular interaction, so that any
difference in outcome is attributable to that term alone. Sections~\ref{sec:flux},
\ref{sec:dynamical}, and~\ref{sec:roaming} locate the three transition states, and
Sec.~\ref{sec:threeradii} relates the tight entropic bottleneck to Makarov's picture.
Section~\ref{sec:conclusions} concludes; computational methods and the gap-time analysis are in
Appendices~\ref{app:methods} and~\ref{app:gap}.

\section{The classical Chesnavich model}
\label{sec:chesnavich}

With zero overall angular momentum and planar geometry, Chesnavich's two-degree-of-freedom (2-DoF) Hamiltonian
is~\cite{chesnavich1986,mauguiere2014jcp,EzraWiggins19}
\begin{equation}
H=\frac{p_r^2}{2\mu}+\frac{p_\theta^2}{2}\!\left(\frac{1}{I_{\mathrm{CH_3}}}+\frac{1}{\mu r^2}\right)+V(r,\theta),
\label{eq:Hches}
\end{equation}
where $r$ is the distance from the $\mathrm{CH}_3^+$ center of mass to the H atom and
$\theta$ the relative orientation. The potential is a barrierless radial channel plus an
orientation-dependent term that hinders free rotation of the fragment, favoring the aligned
geometry $\theta=0$ and switching off with distance:
\begin{align}
V(r,\theta)&=V_{\mathrm{CH}}(r)+\tfrac12 V_0(r)\bigl[1-\cos2\theta\bigr],
\qquad V_0(r)=V_e\,e^{-\alpha(r-r_e)^2},\label{eq:Vches}\\
V_{\mathrm{CH}}(r)&=\frac{D_e}{c_1-6}\Bigl\{2(3-c_2)e^{c_1(1-x)}-(4c_2-c_1c_2+c_1)x^{-6}-(c_1-6)c_2\,x^{-4}\Bigr\},
\label{eq:VCH}
\end{align}
with $x=r/r_e$. As $r$ decreases and $V_0(r)$ grows, this term stiffens and confines the
fragment near alignment; we call it the \emph{orientation lock} (or Chesnavich lock) for
brevity. The parameters, phenomenological rather than fitted to data, are $D_e=47$~kcal/mol, $r_e=1.1~\Aa$, $c_1=7.37$,
$c_2=1.61$, $V_e=55$~kcal/mol, $I_{\mathrm{CH_3}}=2.3734~\mathrm{u\,\Aa^2}$,
$\mu=0.9445~\mathrm{u}$, and $\alpha=1~\Aa^{-2}$~\cite{chesnavich1986,mauguiere2014jcp};
the $x^{-4}$ term is the ion--induced-dipole $-1/r^4$ tail.

The transition states of this surface, which has two degrees of freedom, are periodic orbits, not saddles. As established by
Maugui\`ere, Collins, Ezra, Farantos and Wiggins and by Kraj\v{n}\'ak and Waalkens, the orbits
that organize roaming are unstable periodic orbits on a fixed energy shell, none located at a critical
point of $V(r,\theta)$: an inner orientation-locking (tight) orbit whose \emph{periodic-orbit
dividing surface} (PODS), the dividing surface spanned by an unstable periodic orbit, gates
entry to the well; an outer relative-equilibrium orbit on the centrifugal barrier of the
effective radial potential $V_{\mathrm{CH}}(r)+\tfrac12 p_\theta^2\bigl(1/I_{\mathrm{CH_3}}+1/\mu r^2\bigr)$
(the orbiting, or loose, transition state) whose PODS gates passage to the radical channel; and
an intermediate free-rotor family whose PODS is used to separate direct from roaming
trajectories~\cite{mauguiere2014jcp,KrajnakWaalkens18,KrajnakWiggins18}.

A single quantity captures how tightly the orientation lock---the \emph{angular potential}
$\tfrac12 V_0(r)[1-\cos2\theta]$---confines the fragment at each radius: the \emph{harmonic
transverse frequency}. The transverse degree of freedom is the angular coordinate $\theta$,
transverse to the radial reaction coordinate $r$, and near the aligned geometry $\theta=0$ it
behaves like a vibration. Expanding the angular potential about $\theta=0$
gives $\tfrac12 V_0(r)[1-\cos2\theta]\simeq V_0(r)\,\theta^2$,
a harmonic well whose \emph{curvature}---the second derivative in $\theta$, that is, the angular
force constant---is $2V_0(r)$. The matching kinetic term in Eq.~\eqref{eq:Hches} is
$\tfrac12\bigl(1/I_{\mathrm{CH_3}}+1/\mu r^2\bigr)p_\theta^2$, so the effective angular mass is
$\bigl(1/I_{\mathrm{CH_3}}+1/\mu r^2\bigr)^{-1}$. A harmonic oscillator has frequency
$\sqrt{\text{force constant}/\text{mass}}$, which here is
\begin{equation}
\Omega_{\mathrm{Ch}}(r)=\sqrt{2V_0(r)\Bigl(\tfrac{1}{I_{\mathrm{CH_3}}}+\tfrac{1}{\mu r^2}\Bigr)}.
\label{eq:OmCh}
\end{equation}
It is a local measure of orientational stiffness: the larger it is, the more sharply the fragment
is held near alignment. Why a large transverse frequency should impede the reaction even where the
radial potential is flat is the mechanism we build the next section around; here we note only its
shape. Since $V_0(r)=V_e\,e^{-\alpha(r-r_e)^2}$ falls off monotonically for $r>r_e$, so does
$\Omega_{\mathrm{Ch}}(r)$, which might suggest that the channel has no interior flux minimum.
It does have one. The harmonic frequency omits both
the $r$-dependence of the available channel energy and the anharmonicity of the hindered rotor, and
the exact radial-surface flux computed in Sec.~\ref{sec:flux} has a genuine, if shallow, interior
minimum on the Chesnavich surface. The Chesnavich model therefore does possess an entropic
bottleneck; it is shallow, and arises mainly from the radial variation of channel energy together
with the angular dynamics rather than from a maximum of transverse stiffness.

\section{The designed transverse-stiffness model}
\label{sec:design}

\begin{figure}[t]
\centering
\includegraphics[width=\textwidth]{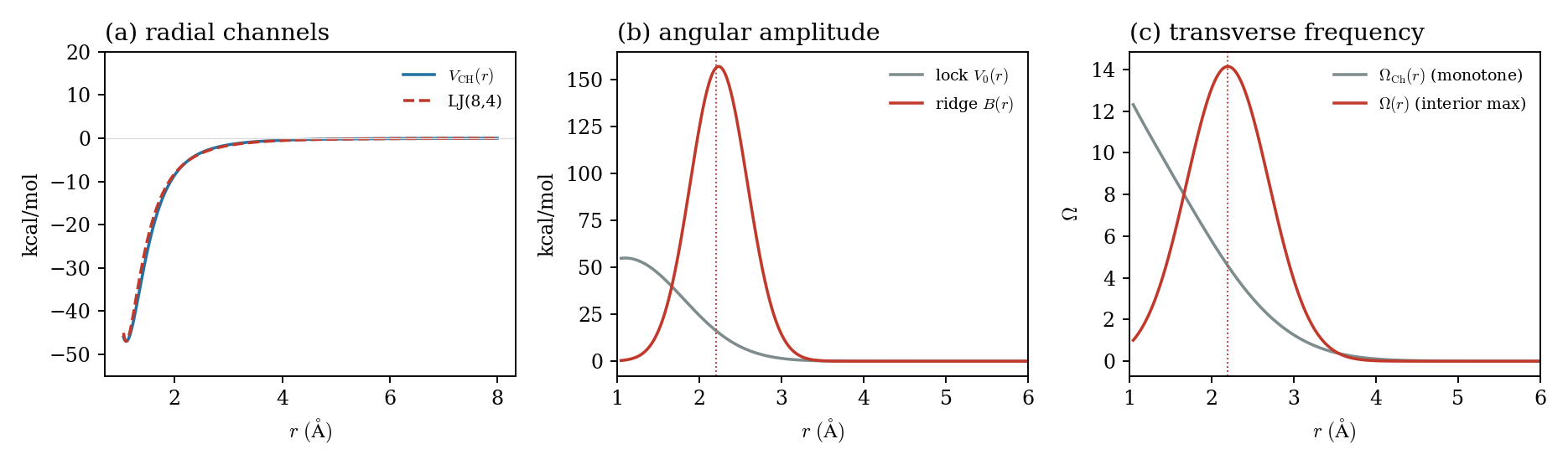}
\caption{Construction of the designed model. (a)~The two barrierless radial channels,
$V_{\mathrm{CH}}(r)$ and the LJ(8,4) form; neither has a radial barrier along the approach, and
both decay as $-1/r^4$ at long range. (b)~Angular hindrance amplitude versus $r$: the monotone
Chesnavich lock $V_0(r)$ ($\alpha=1$) and the entropic ridge $B(r)$, taller and concentrated
near $\rc=2.2~\Aa$. (c)~Harmonic transverse frequency: $\Omega_{\mathrm{Ch}}(r)$ decreases
monotonically, whereas the designed $\Omega(r)$ has a single interior maximum at $\rc$.}
\label{fig:construction}
\end{figure}

We keep the kinetic energy of the Chesnavich model unchanged (the same reduced mass, moment of
inertia, and $r$-dependent angular mass $a(r)$) and keep its long-range attraction (the
ion--induced-dipole $-1/r^4$ tail), but replace the orientation lock by a localized
transverse-stiffness ridge that we \emph{design}, that is, construct by hand to have a prescribed
property rather than fit to a real molecule: a single interior maximum of the transverse
stiffness. The transverse stiffness is the strength of the angular restoring force, measured by
the harmonic transverse frequency $\Omega(r)$; a stiffer transverse mode (larger $\Omega$)
confines the angular motion more tightly. The designed Hamiltonian is
\begin{equation}
H=\frac{p_r^2}{2\mu}+\tfrac12 a(r)\,p_\theta^2+V_{\mathrm{rad}}(r)+\tfrac12 B(r)\bigl[1-\cos2\theta\bigr],
\qquad a(r)=\frac{1}{I_{\mathrm{CH_3}}}+\frac{1}{\mu r^2},
\label{eq:Hdesign}
\end{equation}
in which the angular potential's prefactor $B(r)$---its strength as a function of $r$---is a
localized Gaussian bump centered at an interior radius $\rc$,
\begin{equation}
B(r)=\tfrac12\,a(r)^{-1}\,\Omega_0^2\,e^{-(r-\rc)^2/\sigma^2}.
\label{eq:B}
\end{equation}
By the small-oscillation argument of Sec.~\ref{sec:chesnavich}, now with the ridge amplitude
$B(r)$ in place of the lock amplitude $V_0(r)$ and $a(r)$ the kinetic coefficient of
Eq.~\eqref{eq:Hdesign}, the harmonic transverse frequency about $\theta=0$ is
\begin{equation}
\Omega(r)=\sqrt{2\,B(r)\,a(r)}=\Omega_0\,e^{-(r-\rc)^2/2\sigma^2}.
\label{eq:Omega}
\end{equation}
The factor $a(r)^{-1}$ in $B(r)$ is chosen precisely to cancel the $r$-dependence of the angular
mass, so that $\Omega(r)$ is an exact Gaussian with a single smooth interior maximum at $r=\rc$
(Fig.~\ref{fig:construction}c), in contrast to the monotone $\Omega_{\mathrm{Ch}}(r)$ of the
Chesnavich lock.

Why an interior maximum of $\Omega(r)$ acts as a bottleneck is Makarov's
observation~\cite{makarov2017}, Eq.~\eqref{eq:makarovF}, restated here for the angular mode.
The angular motion is fast compared with the radial approach, so at each $r$ it completes many
oscillations while $r$ barely changes, and one may remove it by averaging. Removing a degree of
freedom in equilibrium statistical mechanics means replacing it by its partition
function---a sum over the states it can occupy---and paying a free-energy cost
$-k_BT\ln(\text{that sum})$. For a harmonic mode of frequency $\Omega(r)$ the classical
transverse partition function, over $\theta$ and $p_\theta$ together, is
$Z_\perp(r)=k_BT/\hbar\Omega(r)$, so the free energy along $r$ becomes
\begin{equation}
G(r)=V_{\mathrm{rad}}(r)-k_BT\ln Z_\perp(r)=V_{\mathrm{rad}}(r)+k_BT\ln\Omega(r)+\text{const},
\label{eq:Gdesign}
\end{equation}
which is Eq.~\eqref{eq:makarovF} with the angular coordinate playing the role of the transverse
mode (the constant absorbs the reference frequency $\omega_0$). The added term is entropic, not
energetic: $\Omega$ enters only through $Z_\perp$, the number of accessible angular states, and
not through the energy of any configuration. A stiffer angular spring (larger $\Omega$) holds the
fragment in a narrower cone about alignment, so fewer orientations are accessible; fewer
accessible states mean lower entropy and therefore higher free energy. At fixed energy the same
statement reads: the number of transverse states below a transverse energy $E_\perp$ scales as
$E_\perp/\Omega$, so a stiffer mode has fewer of them~\cite{makarov2017,truhlar1996}. Wherever
$\Omega(r)$ is largest, $G(r)$ is highest.

The lock and the ridge run through the \emph{same} frequency formula,
$\Omega=\sqrt{2A(r)\,a(r)}$, differing only in the angular amplitude $A(r)$: $A=V_0(r)$ for the
Chesnavich lock and $A=B(r)$ for the designed ridge. For the lock $V_0(r)$ falls off
monotonically, so $\Omega_{\mathrm{Ch}}$ and its free-energy term $k_BT\ln\Omega_{\mathrm{Ch}}$
decrease monotonically with $r$: by this mechanism alone the lock cannot place a bottleneck in the
interior. For the ridge $\Omega(r)$ peaks at $\rc$ [Eq.~\eqref{eq:Omega}], so $k_BT\ln\Omega$
peaks there too---an entropic bottleneck placed in the interior by design, even though
$V_{\mathrm{rad}}(r)$ has no barrier there. This is Makarov's transverse-stiffening mechanism
built into the angular coordinate of a roaming model. The price is a larger angular amplitude:
$B(r)$ reaches $\sim156$~kcal/mol at $\rc$, against $\sim16$~kcal/mol for $V_0(\rc)$
(Fig.~\ref{fig:construction}b).

This canonical $k_BT\ln\Omega$ is a heuristic: we work at fixed energy, not in a thermal bath, so
the quantity we actually compute is its microcanonical counterpart, the exact transverse flux
through a surface at radius $R$, whose interior minimum plays the role of the free-energy maximum
(Sec.~\ref{sec:flux}). The contrast with the lock is one of mechanism, not of outcome: the
Chesnavich model does still have a shallow interior bottleneck, but---as Sec.~\ref{sec:flux}
shows---it comes chiefly from the way the energy available to the angular mode varies with $r$,
together with the full hindered-rotor dynamics, not from a maximum of transverse stiffness. Our
ridge places the bottleneck directly, through the stiffness itself.

For the radial channel we use either the Chesnavich
$V_{\mathrm{CH}}(r)$ or a Lennard-Jones LJ(8,4) form, both decaying as $-1/r^4$ at long range,
\begin{equation}
V_{\mathrm{LJ}}(r)=D_e\Bigl[(r_e/r)^{8}-2(r_e/r)^{4}\Bigr].
\label{eq:lj}
\end{equation}
We use the near-identity of the results under this substitution (Sec.~\ref{sec:matched}) to
establish that the gating is produced by the angular ridge and is insensitive to the shape of the
barrierless radial channel: with the long-range tail and the angular term held fixed, replacing
$V_{\mathrm{CH}}(r)$ by $V_{\mathrm{LJ}}(r)$ leaves the inner-capture drop essentially unchanged.
The ridge itself is radially localized at $\rc$; what the substitution rules out is a dependence
on the detailed shape of the radial channel potential, not radial structure as such. The model
parameters are listed in Table~\ref{tab:param}.

\begin{table}[t]
\centering
\caption{Parameters of the designed model, Eqs.~\eqref{eq:Hdesign}--\eqref{eq:lj}. The
``sharp ridge'' values are used throughout unless stated otherwise. Energies in kcal/mol,
lengths in \AA, masses in u.}
\label{tab:param}
\begin{tabular}{lll}
\toprule
quantity & symbol & value \\
\midrule
reduced mass & $\mu$ & $0.9445$ \\
$\mathrm{CH_3}$ moment of inertia & $I_{\mathrm{CH_3}}$ & $2.3734$ \\
well depth / tail scale & $D_e,\ r_e$ & $47,\ 1.1$ \\
Chesnavich shape constants & $c_1,c_2$ & $7.37,\ 1.61$ \\
Chesnavich lock height / width & $V_e,\ \alpha$ & $55,\ 1$ \\
ridge peak frequency squared & $\Omega_0^2$ & $200$ \\
ridge center / width & $\rc,\ \sigma$ & $2.2,\ 0.5$ \\
roaming energy & $E$ & $2$ \\
\bottomrule
\end{tabular}
\end{table}

\section{The entropic ridge gates inner capture and displaces roaming outward}
\label{sec:matched}

The model was built to ask a dynamical question, and we answer it first, before locating the
phase-space structures that explain the answer. The question is what the localized entropic
ridge does to roaming \emph{relative to} the orientation-locking interaction of the original
Chesnavich surface. We answer it with a controlled comparison in which everything but the
angular interaction is held fixed. We fix the radial channel at the Chesnavich potential
$V_{\mathrm{CH}}(r)$, fix the energy $E=2$, and fix the incoming fixed-energy ensemble; the
only difference between the two runs is the angular term, the Chesnavich lock
$\tfrac12 V_0(r)[1-\cos2\theta]$ in one and the entropic ridge $\tfrac12 B(r)[1-\cos2\theta]$ in
the other. The ensemble is a $240\times240$ grid in $(\theta_0,p_{\theta,0})$ launched inward
from $r_0=6~\Aa$ at fixed $E$, with the inward radial momentum fixed by the energy constraint
and energetically inaccessible launch points removed; a trajectory is counted as reaching the
well if it crosses $r<1.6~\Aa$, and the integrator and ensemble are documented in
Appendix~\ref{app:methods}. This is the dynamical counterpart of the matched control at the
level of the flux (Sec.~\ref{sec:flux}), where the same ridge is placed on $V_{\mathrm{CH}}(r)$
at the level of the exact directional flux; here we compare its trajectory consequences against
the lock at fixed radial energetics.

\begin{figure}[t]
\centering
\includegraphics[width=\textwidth]{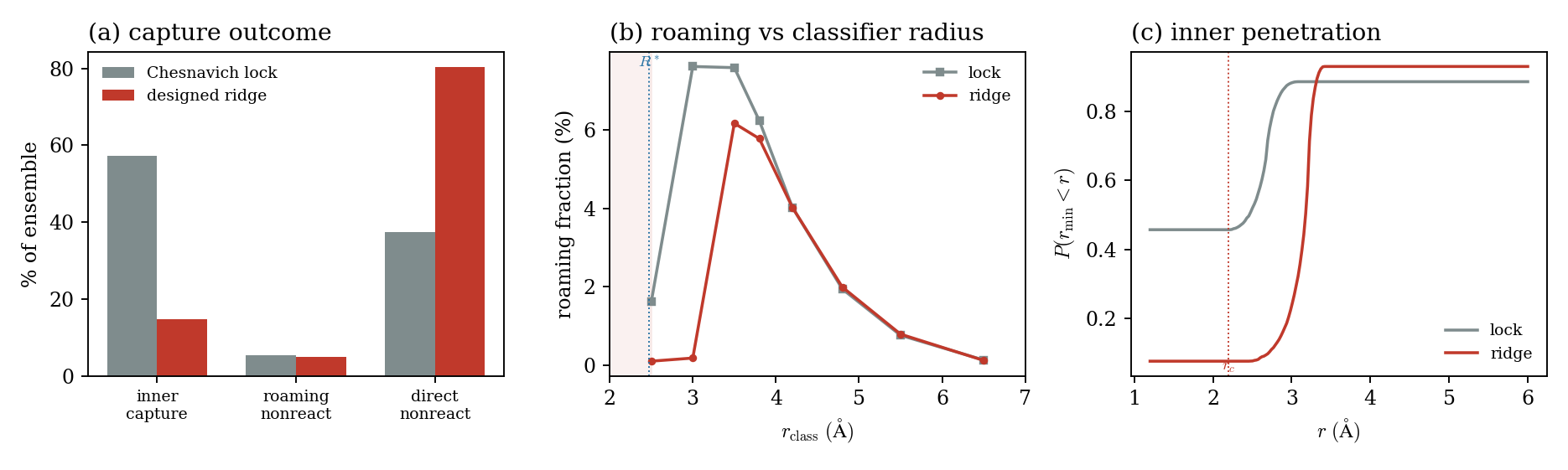}
\caption{The controlled comparison at $E=2$. The radial channel $V_{\mathrm{CH}}(r)$, the
energy, and the inward fixed-energy ensemble are identical in the two runs; only the angular
term differs, set to the Chesnavich lock (the orientation-locking term of the original model)
in one and to the entropic ridge (the localized transverse-stiffness maximum at $\rc$) in the
other. (a)~Inner-capture outcome: the ensemble partitioned into inner capture (entry to
$r<1.6~\Aa$), roaming nonreactive, and direct nonreactive. The ridge cuts inner capture from
$57.3\%$ to $14.7\%$ and raises direct nonreactive return from $37.4\%$ to $80.3\%$, while the
roaming (nonreactive) fraction is nearly unchanged ($5.3\%\to4.9\%$). (b)~The roaming fraction versus classifier radius: under the ridge it is suppressed inside $\rc$,
switches on beyond it, and persists farther out than under the lock, crossing above the lock's
curve beyond $\sim3.8~\Aa$. The variational bottleneck $\Rstar=2.47~\Aa$ for this channel is
marked. (c)~Cumulative inner penetration $P(r_{\min}<r)$: the lock funnels trajectories into the
well; the ridge gates them just outside $\rc$.}
\label{fig:matched}
\end{figure}

\paragraph{What changes.}
Figure~\ref{fig:matched} reports the comparison, and two effects are large and unambiguous.
First, the ridge \emph{gates entry into the inner well}: entry into the well
($r<1.6~\Aa$) falls from $57.3\%$ under the monotone lock to $14.7\%$ under the ridge, and the
fraction of the ensemble returned directly to the outer region without entering the well rises
correspondingly from $37.4\%$ to $80.3\%$ (Fig.~\ref{fig:matched}a). The mechanism is visible
in the penetration profile (Fig.~\ref{fig:matched}c): the monotone lock funnels incoming
trajectories toward $\theta=0$ and into the well, so the cumulative penetration
$P(r_{\min}<r)$ rises steadily; the ridge instead reflects all but a narrow channel near
$\theta=0$, and $P(r_{\min}<r)$ is gated sharply just outside $\rc$. Second, the ridge
\emph{does not increase the roaming fraction}; it \emph{displaces the roaming fraction outward}, where the roaming fraction counts
trajectories that cross the classifier radius three or more times, in either direction, before
reacting or escaping. At the
classifier radius $r_{\mathrm{class}}=3.5~\Aa$ the roaming (nonreactive) fraction is
essentially unchanged, $5.3\%\to4.9\%$. But the radial profile of the statistic changes
qualitatively (Fig.~\ref{fig:matched}b): under the monotone lock repeated crossings are already
present at smaller radii and decay quickly outward; under the ridge they are suppressed inside
$\rc$, switch on sharply just beyond it, and persist to larger radii, exceeding the lock's curve
at every radius beyond $\sim3.8~\Aa$. The two runs agree closely when the LJ(8,4) channel is
substituted for $V_{\mathrm{CH}}(r)$ (inner capture $57.3\%/15.1\%$, within $1\%$), confirming
that the effect is carried by the angular interaction and not by the radial form.

\paragraph{What this is, and what it is not.}
The result carries four qualifications.
\emph{(i)~``Reactive'' means entry into the well, not committed product.}
A trajectory is counted as reactive if its closest approach reaches $r<1.6~\Aa$. The flow is
volume-preserving and the well is not an attractor, so this denotes entry into the molecular
well, not a committed product yield: a captured trajectory may recross and redissociate. The
gating result is a statement about inner-well capture (Appendix~\ref{app:methods}), and the same
definition is used for the classifier ensemble of Sec.~\ref{sec:roaming}.
\emph{(ii)~Amplitude and localization, separated by a control.}
The two angular interactions differ in \emph{amplitude} as well as in radial profile: the
ridge reaches $\sim156$~kcal/mol at $\rc$, against $\sim16$~kcal/mol for the lock at the same
radius. To separate the two effects we ran strength-matched monotone controls: the same lock
form $V_e\,e^{-(r-r_e)^2}$ with $V_e$ raised, on the same channel and ensemble. Capture falls
monotonically with the hindrance strength at the bottleneck radius, $V_0(\rc)$: raising it from
the standard $16.4$ to $46.5$, $80$, and $156$~kcal/mol gives capture $57.3\%$, $33.0\%$,
$24.9\%$, and $17.8\%$. The $\rc$-matched monotone lock ($V_0(\rc)=156$, requiring $V_e=523$)
also displaces roaming outward much as the ridge does (switching on near $3.8~\Aa$), and its
capture transfers between radial channels ($17.8\%$ on both $V_{\mathrm{CH}}$ and LJ). Both
headline effects therefore follow from strong orientational hindrance in the region of the
bottleneck radius and are not unique to an interior maximum. What the interior maximum supplies
is placement and economy: it concentrates the hindrance at the gating radius, so the ridge gates
far more effectively than a monotone profile of equal global peak strength ($14.7\%$ against
$33.0\%$) and does so without imposing a very large hindrance at all smaller radii; the residual
difference between the ridge and the $\rc$-matched control ($14.7\%$ against $17.8\%$) is the
contribution of the localized profile itself.
\emph{(iii)~The roaming fraction is parameter-dependent; the redistribution is not.}
The roaming fraction is sensitive to the Chesnavich switching parameter $\alpha$. For the
early-switching lock ($\alpha=4$), whose hindrance has all but vanished by $r\simeq2~\Aa$,
almost the entire incoming ensemble is captured ($\sim99.9\%$) and the baseline roaming
fraction is tiny ($\sim0.2\%$); against that baseline the ridge both reduces capture and
\emph{increases} the measured roaming fraction. It would therefore be misleading to speak of
``the'' roaming fraction of the Chesnavich model. The robust result, common to $\alpha=1$ and
$\alpha=4$, is the inner-well gating; the outward redistribution is demonstrated explicitly
for the $\alpha=1$ comparison through the full classifier-radius sweep of
Fig.~\ref{fig:matched}b, while the raw roaming percentage at any single classifier radius
changes strongly with $\alpha$.
\emph{(iv)~This is conditional-on-entry intermediate dynamics.}
The launch surface $r_0=6~\Aa$ lies \emph{inside} the outer orbiting transition state
($r_{\mathrm{OTS}}\simeq7.48~\Aa$, Sec.~\ref{sec:dynamical}); this is a comparison of
intermediate-region dynamics conditional on inward entry, not a complete reactant-to-product
scattering calculation.

The remaining sections locate the phase-space structures that explain this behavior: a
variational flux bottleneck just outside the ridge, an invariant dynamical transition state far
outside it, and a classifier between the two that tracks neither.

\section{The variational flux bottleneck}
\label{sec:flux}

The statistical bottleneck is the dividing surface spanned by the tight orbit. For a radial dividing surface $r=R$ at energy
$E$, the exact fixed-energy directional flux is~\cite{wsw2008}
\begin{equation}
\Phi_R(E)=\oint_{0}^{2\pi}\! 2\sqrt{\frac{2\,[\,\varepsilon(R)-U_{\mathrm{ang}}(R,\theta)\,]_+}{a(R)}}\,d\theta,
\qquad \varepsilon(R)=E-V_{\mathrm{rad}}(R),
\label{eq:flux}
\end{equation}
with $U_{\mathrm{ang}}(R,\theta)=\tfrac12 B(R)[1-\cos2\theta]$ and $[\cdot]_+$ the positive
part; $\Phi_R(E)$ is the area of the transverse (angular) phase space $(\theta,p_\theta)$, on
the energy shell at $r=R$, that is energetically able to cross outward. The variational
bottleneck $\Rstar$ minimizes $\Phi_R(E)$ over $R$~\cite{truhlar1996,pollakpechukas1978,pechukaspollak1979};
we work at fixed energy throughout, and $\Rstar$ is the fixed-energy analogue of the maximum of
the canonical constrained free energy of Eq.~\eqref{eq:makarovF}, which is why we describe it as
entropic. We also define the angular confinement factor
\begin{equation}
C(R,E)=\frac{\Phi_R(E)}{\Phi_R^{\mathrm{free}}(E)},
\qquad \Phi_R^{\mathrm{free}}(E)=2\pi\cdot 2\sqrt{2\,\varepsilon(R)/a(R)},
\label{eq:Cfac}
\end{equation}
the fraction of the free-rotor transverse phase space that remains accessible; normalizing a
hindered quantity by its free-rotor value follows the hindered-rotor treatment of Pitzer and
Gwinn~\cite{pitzergwinn1942}.

\begin{figure}[t]
\centering
\includegraphics[width=\textwidth]{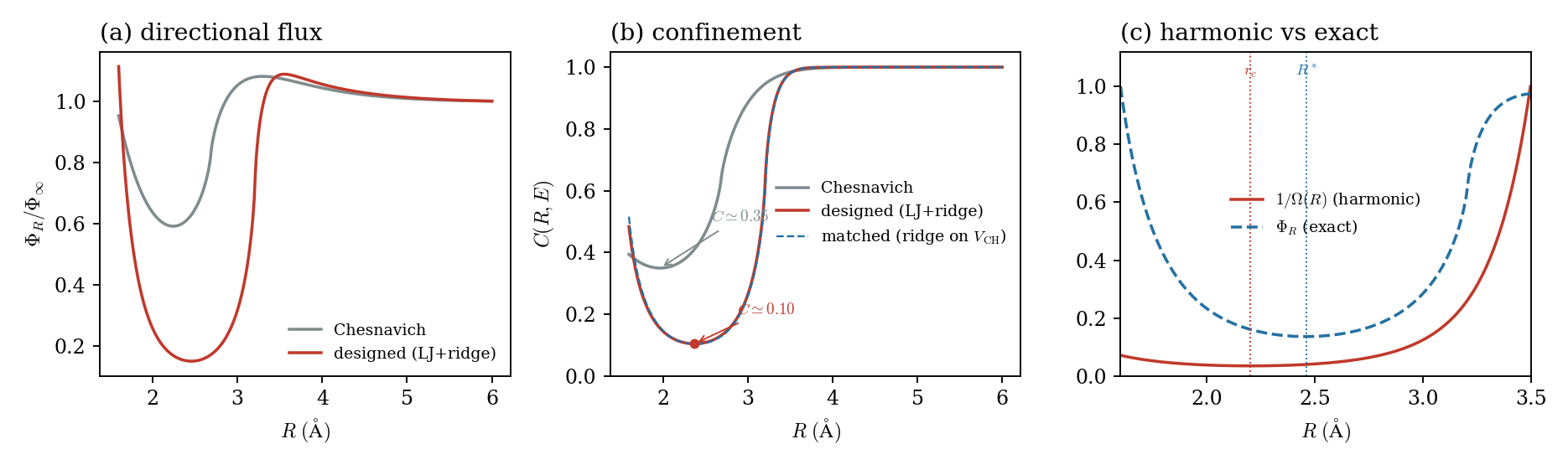}
\caption{The variational flux bottleneck at $E=2$. (a)~Exact directional flux
$\Phi_R/\Phi_\infty$, Eq.~\eqref{eq:flux}, for the designed (LJ+ridge) and Chesnavich
surfaces; the designed model has a deep, narrow minimum where the Chesnavich surface has only
a shallow one. (b)~Angular confinement $C(R,E)$, Eq.~\eqref{eq:Cfac}: the designed
constriction ($C\simeq0.10$) is reproduced by the matched control (ridge on the Chesnavich
channel, dashed), well below the Chesnavich value ($C\simeq0.35$). (c)~Decomposition of the
designed-model flux: the harmonic approximation $1/\Omega$ is minimized at $\rc=2.20~\Aa$, but
the exact flux minimum sits at $\Rstar=2.46~\Aa$ because the available channel energy
$\varepsilon(R)/\Omega$ shifts it outward.}
\label{fig:flux}
\end{figure}

Figure~\ref{fig:flux} reports these quantities at $E=2$. The designed model has a deep, narrow
flux minimum at $\Rstar\simeq2.46~\Aa$ (designed channel), with $\Phi_{R^*}/\Phi_\infty\simeq
0.15$ and confinement $C\simeq0.10$: a deep entropic bottleneck on an otherwise barrierless
channel, in the sense that only about a tenth of the free transverse phase space still crosses
($C\simeq0.10$). The Chesnavich surface, by contrast, has only a shallow interior minimum at
$\Rstar\simeq2.24~\Aa$ with $C\simeq0.35$, confirming the point of Sec.~\ref{sec:chesnavich}
that the Chesnavich channel \emph{does} possess an entropic bottleneck, but a weak one. The
matched control at the level of the flux (the ridge placed on the Chesnavich channel
$V_{\mathrm{CH}}(r)$) reproduces the deep constriction almost exactly ($\Rstar\simeq2.47~\Aa$,
$C\simeq0.10$; dashed curve in Fig.~\ref{fig:flux}b), demonstrating that the deep statistical
bottleneck is produced by the designed transverse stiffness and transfers between radial
channels. The same designed transverse stiffness produces both this deep flux minimum and the
dynamical gating of Sec.~\ref{sec:matched}, at their respective radii.

Two features deserve emphasis. First, the flux minimum is genuinely entropic: there is no
radial potential barrier along the channel (Fig.~\ref{fig:construction}a), so the constriction
is entirely an effect of angular confinement, the reduction of the accessible transverse phase
space to $C\simeq0.10$, not of any radial potential barrier. Second, the flux minimum $\Rstar\simeq2.46~\Aa$ does \emph{not} coincide with the maximum of the
designed transverse stiffness at $\rc=2.20~\Aa$; it sits a little outside it. The decomposition in
Fig.~\ref{fig:flux}c shows why: the harmonic approximation $1/\Omega(R)$ is minimized at $\rc$, but
the exact flux carries the additional factor of available channel energy $\varepsilon(R)$, which
rises with $R$ along the attractive channel and shifts the true minimum outward. The surface that
actually constricts the flux is therefore this radius, a little outside the stiffness maximum.

\section{The dynamical transition-state structure}
\label{sec:dynamical}

\begin{figure}[t]
\centering
\includegraphics[width=\textwidth]{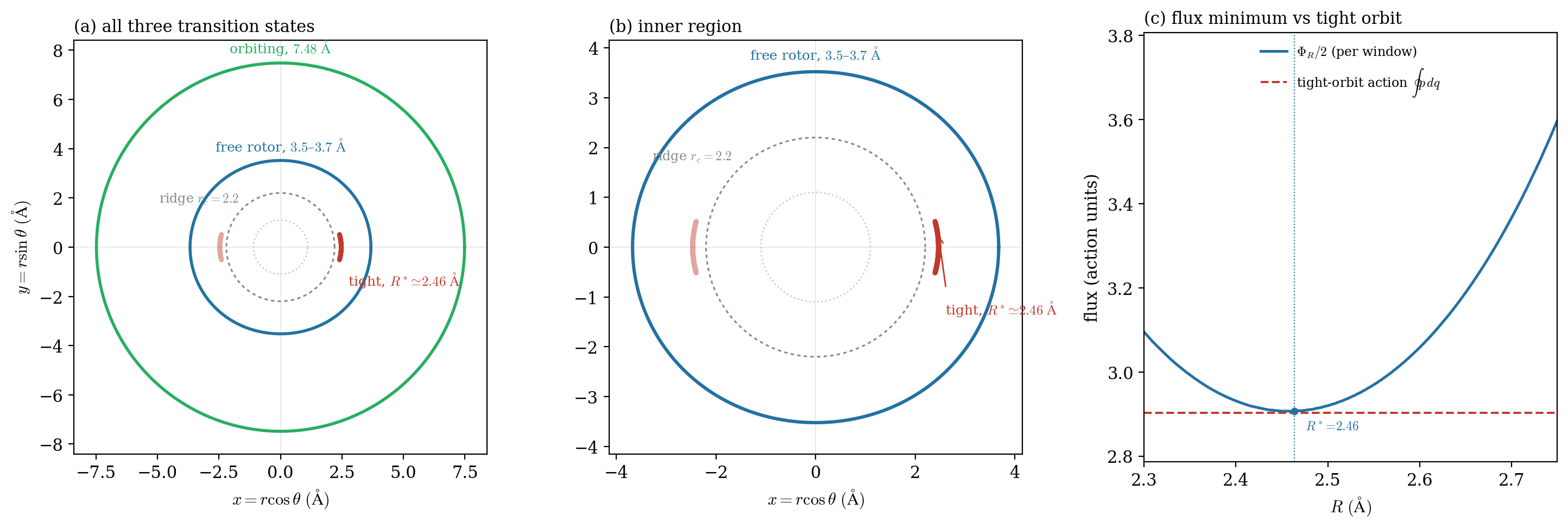}
\caption{The three transition states of the designed model at $E=2$, all unstable periodic orbits,
in the configuration plane $(x,y)=(r\cos\theta,r\sin\theta)$. (a)~All three: the orbiting orbit
(circle, $r_{\mathrm{OTS}}=7.48~\Aa$), the free-rotor orbit (loop, $r\in[3.5,3.7]~\Aa$), and the
tight orbit (short arc near the aligned axis, $\Rstar\simeq2.46~\Aa$; its $\theta=\pi$ mirror is
shown faintly). (b)~The inner region: the tight orbit is a libration of $\pm12^\circ$ about the
aligned axis at the flux minimum $\Rstar$, where the radial channel is barrierless; the free-rotor
orbit is a rotation. The ridge center $\rc=2.2~\Aa$ is marked in both panels. (c)~The
fixed-radius directional flux per aligned window, $\Phi_R/2$, against the action
$\oint p\,dq=2.902$ of the tight orbit (dashed line): the minimum, $2.907$ at $R^*$, agrees with
the orbit's dividing-surface flux to $0.15\%$.}
\label{fig:orbits}
\end{figure}

The designed model's three transition states are all unstable periodic orbits, located at $E=2$
and shown in Fig.~\ref{fig:orbits}: the tight orbit at $\Rstar\simeq2.46~\Aa$, a libration of
$\pm12^\circ$ about the aligned axis, at the flux-minimum radius of Sec.~\ref{sec:flux}; the
free-rotor orbit over $3.5$--$3.7~\Aa$, used to classify roaming in Sec.~\ref{sec:roaming}; and the
outer orbiting orbit at $r_{\mathrm{OTS}}=7.48~\Aa$. This section locates the last of these.

\begin{figure}[t]
\centering
\includegraphics[width=0.62\textwidth]{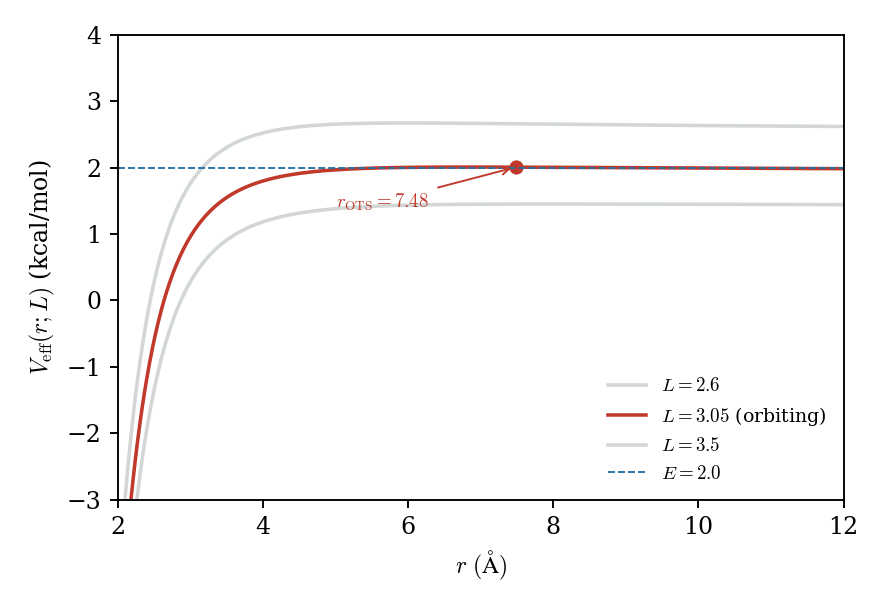}
\caption{The outer orbiting transition state of the designed model (LJ(8,4) channel).
Outer-region effective radial potential $V_{\mathrm{eff}}(r;L)$ at $E=2$: at $L=3.047$ the
orbiting transition state is the unstable centrifugal barrier maximum at
$r_{\mathrm{OTS}}=7.48~\Aa$. The ridge is negligible in this region, so the orbiting transition
state is the standard centrifugal (Langevin) barrier.}
\label{fig:outerTS}
\end{figure}

The invariant phase-space object is the outer orbiting transition state that
gates escape. The long-range tail supports an orbiting relative equilibrium on the centrifugal
barrier of the rotationally averaged effective radial potential
$V_{\mathrm{eff}}(r;L)=V_{\mathrm{rad}}(r)+\tfrac12 L^2 a(r)+\tfrac12 B(r)$.
Figure~\ref{fig:outerTS} shows that at $E=2$ this potential has a barrier maximum at
$r_{\mathrm{OTS}}\simeq7.48~\Aa$ with angular momentum $L=p_\theta=3.047$, the internal rotational momentum (the total angular momentum is zero); the maximum has negative
radial curvature, so the orbit is unstable. The full model is not globally rotationally
symmetric when $B\neq0$, so rotational averaging does not in general define an exact orbit; here,
however, the ridge is negligible at this radius ($B(r_{\mathrm{OTS}})\sim10^{-47}$), so
rotational averaging is exact to numerical precision and this is the asymptotic orbiting relative
equilibrium. The orbiting transition state is the standard centrifugal (Langevin) barrier,
produced by the centrifugal term $\tfrac12 L^2 a(r)$ against the $-1/r^4$ attraction with no
contribution from the ridge. The dividing surface attached to it is defined at this energy and
angular momentum ($E=2$, $L=3.047$); it is not a fixed radius crossed by every trajectory, since
the orbiting-barrier location depends on angular momentum. We do not launch trajectories from
this surface: the inward ensemble of Sec.~\ref{sec:roaming} starts at $r_0=6~\Aa$, inside
$r_{\mathrm{OTS}}$, and its largest angular momentum equals $L=3.047$, so the launch grid is
bounded in angular momentum by the orbiting transition state while lying radially inside it.

This is the second of the three radii. The dynamical transition state that gates escape, the
orbiting orbit at $r_{\mathrm{OTS}}\simeq7.48~\Aa$, lies well outside the variational flux
bottleneck $\Rstar\simeq2.46~\Aa$. These are different physical objects, an inner entropic
constriction and an outer centrifugal escape barrier, so their separation is expected; the
substantive point, completed in Sec.~\ref{sec:threeradii}, is that neither coincides with the
classifier, and no single radius summarizes the dynamics.

\section{The roaming classifier}
\label{sec:roaming}

Maugui\`ere, Collins, Ezra, Farantos and Wiggins constructed phase-space dividing surfaces from
the periodic orbits of the Chesnavich model and classified trajectories by how many times they
cross them~\cite{mauguiere2014jcp}. Kraj\v{n}\'ak and Waalkens computed the stable and unstable
manifolds of those orbits and described the transport through their lobes~\cite{KrajnakWaalkens18};
Kraj\v{n}\'ak and Wiggins obtained the same structures with Lagrangian descriptors, computationally
simpler than the manifold construction~\cite{KrajnakWiggins18}. We adopt their distinction between
direct and roaming trajectories. In the phase-space picture the surface that draws this distinction
is the free-rotor periodic-orbit dividing surface; in the designed model at $E=2$ this is a genuine
hyperbolic orbit, a rotational orbit oscillating radially over $r\in[3.52,3.68]~\Aa$, and we count
crossings of a surface of fixed radius $r_{\mathrm{class}}=3.5~\Aa$, at its inner turning point, as
a practical stand-in.

For the classifier analysis below we use the LJ(8,4) radial channel, for which the inner-capture
fractions quoted in Sec.~\ref{sec:matched} are $57.3\%$ and $15.1\%$. We integrate the inward
ensemble of Sec.~\ref{sec:matched}, here on this channel, and classify each trajectory two ways.
A trajectory is \emph{reactive} if its closest approach reaches the well ($r<1.6~\Aa$) and
\emph{nonreactive} otherwise; it is \emph{roaming} if it crosses $r_{\mathrm{class}}$ three or more
times, in either direction, before escaping, and \emph{direct} otherwise. This differs from the
original scheme in one respect: there the crossing count alone carries both roaming and reactivity,
through the opposite crossing parity of trajectories ending inside versus outside; we read
reactivity directly from the closest approach and use a single crossing threshold for roaming. The
four combinations are direct-reactive, direct-nonreactive, roaming-reactive, and roaming-nonreactive.
Trajectories are integrated to escape or to the time limit, and are not stopped at the well; the two
roaming classes therefore come from the same calculation.

Passing both angular interactions through this classifier on the same radial channel isolates what
the ridge does. Under the ridge the ensemble divides into $14.9\%$ direct-reactive, $78.9\%$
direct-nonreactive, $6.0\%$ roaming-nonreactive, and $0.13\%$ roaming-reactive; under the lock, into
$55.7\%$, $36.8\%$, $6.0\%$, and $1.6\%$. The total reactive fraction, $57.3\%$ for the lock and
$15.1\%$ for the ridge, reproduces the LJ-channel inner-capture numbers quoted in Sec.~\ref{sec:matched}, since
reaching the well is the event scored there; the classifier and the capture partition are two
readings of one calculation. The roaming fraction is comparable between the two models ($7.6\%$ for
the lock, $6.2\%$ for the ridge at $3.5~\Aa$), but its radial location is not: under the lock roaming
already appears at smaller radii ($1.6\%$ at $2.5~\Aa$, rising to $7.6\%$ by $3.0~\Aa$), whereas under
the ridge it is negligible inside $3.0~\Aa$ (below $0.2\%$) and switches on sharply near $3.5~\Aa$
(Fig.~\ref{fig:matched}b). The ridge does not suppress roaming; it displaces it outward, to just
beyond the stiffness maximum.

Whether the intermediate dynamics is statistical is tested by the gap time, the residence time
between entering and leaving the interaction region: a statistical (Rice--Ramsperger--Kassel--Marcus, RRKM) population decays as a
single exponential, so any departure from one shows that a single rate through the bottleneck does
not capture the dynamics. The distribution is non-exponential for both the lock and the ridge
(Appendix~\ref{app:gap}), consistent with the nonstatistical roaming region found by Maugui\`ere,
Collins, Ezra, Farantos and Wiggins~\cite{mauguiere2014jcp,thiele1962,ezra2009gap}.

\begin{figure}[t]
\centering
\includegraphics[width=\textwidth]{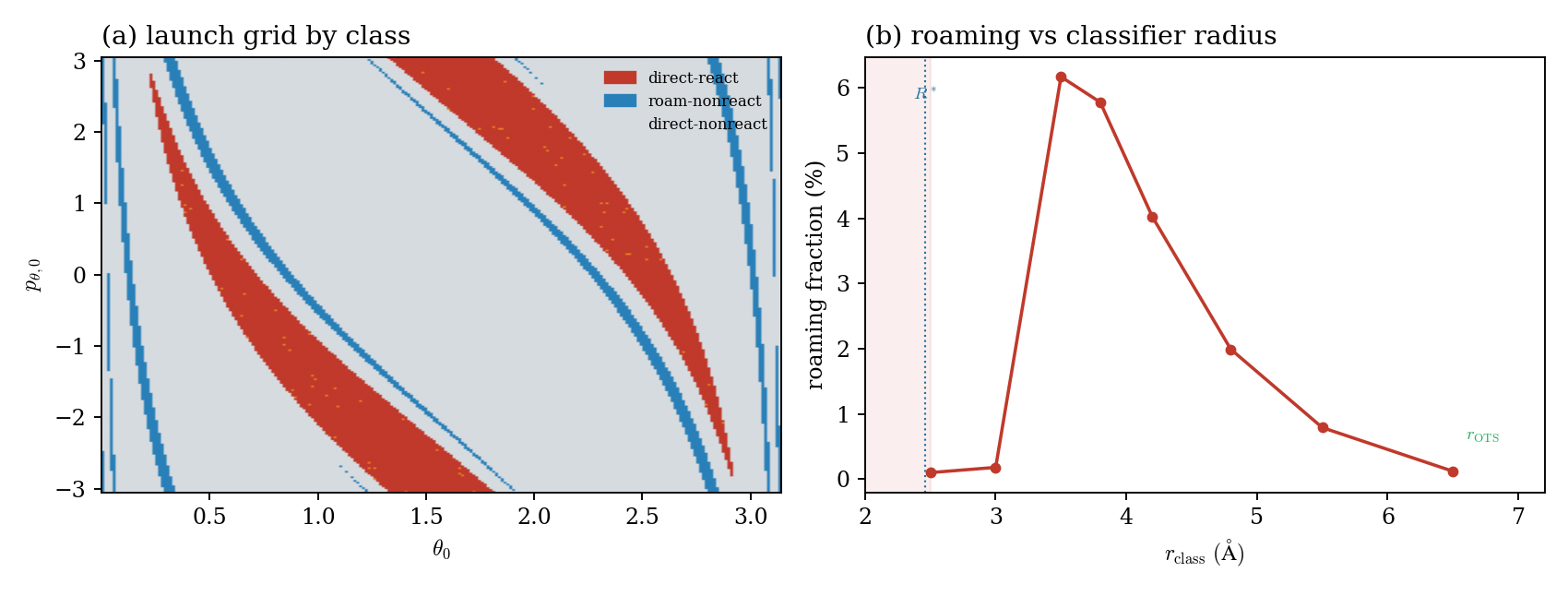}
\caption{Roaming on the designed surface at $E=2$ (LJ(8,4) channel; $240\times240$ inward grid at
$r_0=6~\Aa$). (a)~Launch grid $(\theta_0,p_{\theta,0})$ colored by trajectory class: reactive
trajectories form centrifugal-trapping bands, roaming sets are thin, and most trajectories are
direct and nonreactive. The classification is that of the non-absorbing run, in which a small
roaming-reactive class appears. (b)~Roaming fraction versus classifier radius $r_{\mathrm{class}}$:
suppressed inside the ridge (shaded), switching on near $3.5~\Aa$, and decaying toward zero past the
orbiting radius $r_{\mathrm{OTS}}=7.48~\Aa$. The flux minimum $\Rstar=2.46~\Aa$ and the orbiting
radius are marked; the classifier radius is where this statistic is maximal, not a periodic-orbit
dividing surface.}
\label{fig:roaming}
\end{figure}

The classifier radius is not sharply determined: the roaming fraction against $r_{\mathrm{class}}$
has a broad maximum near $3.4$--$3.5~\Aa$, so radii across that range classify almost identically, and
we take $3.5~\Aa$ as representative. Figure~\ref{fig:roaming}b shows the corresponding LJ(8,4)-channel roaming fraction over the full
radial range, together with the flux minimum $\Rstar$, the ridge
region, and the orbiting radius $r_{\mathrm{OTS}}$, to place the classifier radius among the
transition-state radii. Figure~\ref{fig:roaming}a shows the launch grid colored by class. The reactive trajectories fall into bands, centrifugal-trapping
bands separated by thin roaming sets, whose arrangement can be fractal~\cite{griceandrews1987}. A
separate inner feature near $2~\Aa$ comes from trajectories oscillating in the inner region, inside
the ridge but above $1.6~\Aa$, so never reaching the well, and is not outer roaming. Toward large
radius the roaming fraction dies out as trajectories escape past the orbiting orbit at
$r_{\mathrm{OTS}}=7.48~\Aa$.

\section{The tight transition state}
\label{sec:threeradii}

The designed model carries analogues of the three transition states that organize the Chesnavich model, all
of them unstable periodic orbits located here at $E=2$: the tight orbit at $r\simeq2.46~\Aa$, the
free-rotor orbit spanning $3.52$--$3.68~\Aa$, and the orbiting orbit at $r_{\mathrm{OTS}}=7.48~\Aa$,
introduced in Sec.~\ref{sec:intro}. The tight orbit gates entry to the well and
is the tight transition state of the original model, present here in the same kind: a hindered-rotor
orbit librating within $\pm12^\circ$ of the aligned axis (Fig.~\ref{fig:orbits}). Chesnavich's
original study concerned exactly the competition among such transition states; here we ask what has
changed about the innermost one.

Both models place the tight transition state at a flux minimum where the radial channel is
barrierless: the crossing is constricted not by a potential maximum along $r$ but because angular
hindrance narrows the accessible transverse phase space. In this sense the tight transition state is
entropic in the Chesnavich model as much as in the designed model, and Chesnavich's transition-state
switching is a competition between entropic bottlenecks, neither anchored to a potential saddle. The
flux-minimizing dividing surface is moreover the surface spanned by the tight periodic orbit, since the minimum-flux dividing
surface is bounded by an unstable periodic orbit~\cite{pollakpechukas1978}: the variational
bottleneck $\Rstar\simeq2.46~\Aa$ of Sec.~\ref{sec:flux} sits at the tight orbit. We verify
this directly rather than assume it: in two degrees of freedom the flux through the dividing
surface spanned by a periodic orbit equals the orbit's action $\oint p\,dq$~\cite{pollakpechukas1978,
wsw2008}; for the numerically located tight orbit this action is $2.902$, and the
fixed-radius flux minimum per aligned window is $\Phi_{R^*}/2=2.907$, an agreement of $0.15\%$
(Fig.~\ref{fig:orbits}c; the factor of two counts the two aligned windows $\theta=0,\pi$, each
spanned by one tight orbit). The fixed-radius family thus attains the periodic-orbit flux, and the
identification of the variational bottleneck with the tight orbit's dividing surface is a computed
result, not an assumption. What the
designed model changes is the depth and location of this entropic bottleneck. The interior maximum of
$\Omega$ places a deep constriction at an interior radius ($\Rstar\simeq2.46~\Aa$, $C\simeq0.10$),
where the monotone stiffness of the Chesnavich lock, largest at the inner edge and falling outward,
gives only a shallow one ($C\simeq0.35$, Sec.~\ref{sec:flux}). It is this deep, interior entropic
bottleneck, not the entropic character as such, that gates inner capture and displaces roaming
outward (Secs.~\ref{sec:matched},~\ref{sec:roaming}).

A statistical rate theory would place the bottleneck at $\Rstar$ and compute a rate through it. But
the tight transition state gates only entry; the trajectories it admits do not settle into a single
complex, they roam, sorted by the free-rotor surface and released past the orbiting surface, with
non-exponential gap-time distributions (Sec.~\ref{sec:roaming}). The entropic bottleneck locates
where capture is controlled; it does not by itself describe the roaming and escape that follow. That
the controlling bottleneck is entropic rather than energetic is the point of contact with Makarov.

\paragraph{Relation to Makarov.}
Makarov~\cite{makarov2017} considered a reaction coordinate crossing a region where a transverse
mode stiffens, with no potential barrier along the coordinate; integrating out the transverse mode
yields a free-energy barrier $G(x)=V(x)+k_BT\ln[\omega_\perp(x)/\omega_0]$ located at the stiffness
maximum, a purely entropic barrier. Our fixed-energy $\Omega(r)$ is the microcanonical counterpart of
his $\omega_\perp(x)$, and Eq.~\eqref{eq:makarovF} is the same relation. Makarov further
observed that at such a barrier different observables need not agree: the reaction rate is well
described by transition-state theory while the distribution of transition-path times is not, because
the slow transverse mode carries dynamics that a single rate does not capture. In his model the
entropic maximum, the geometric constriction, and the rate-controlling surface all sit at one
location. Here the tight transition state is likewise entropic, but it is one of several transition
states at separated radii, and the roaming it admits is the nonstatistical dynamics that a rate
through it does not describe: a Hamiltonian roaming realization of the same rate-versus-dynamics
distinction.

\section{Conclusions}
\label{sec:conclusions}

We constructed a designed Chesnavich-type roaming model in which the orientation-locking angular
hindrance is replaced by a localized maximum of the transverse frequency, a transverse-stiffness
ridge on an otherwise barrierless channel, and compared two fixed-energy trajectory ensembles that
differ only in the angular interaction, the Chesnavich lock in one and the entropic ridge in the
other, with the same radial channel, energy, and inward initial conditions. The central result is
dynamical: with the same radial channel, energy, and inward fixed-energy ensemble, the entropic
ridge gates entry into the inner well and converts most of the incoming flux into direct nonreactive
return; it does not increase the roaming fraction at the classifier radius but spatially reorganizes
it, suppressing it inside the ridge and displacing it outward relative to the Chesnavich lock.
Inner-well capture falls from $57.3\%$ to $14.7\%$ and direct nonreactive return rises from $37.4\%$
to $80.3\%$, while the nonreactive roaming fraction at $3.5~\Aa$ is essentially unchanged
($5.3\%\to4.9\%$) but is pushed outward in radius. The gating is robust across the Chesnavich
switching parameter $\alpha$, and is essentially
unchanged when the Chesnavich radial potential is replaced by an LJ(8,4) channel with the same
$-1/r^4$ tail; the outward displacement is established for the $\alpha=1$ comparison.
Strength-matched monotone controls separate amplitude from localization: capture tracks the
hindrance strength at the bottleneck radius, and a monotone lock matched to the ridge there
reproduces both the gating ($17.8\%$ capture) and the outward displacement, while at equal global peak
amplitude the ridge gates far more effectively than a monotone profile ($14.7\%$ against
$33.0\%$). The interior maximum supplies placement, concentrating the hindrance strength at the
gating radius.

The phase-space explanation is the set of unstable periodic orbits and their dividing surfaces that
control capture and roaming: the tight orbit that gates the well, the free-rotor orbit that sorts
direct from roaming, and the orbiting orbit that gates escape, at $\Rstar\simeq2.46~\Aa$,
$r_{\mathrm{class}}\simeq3.5~\Aa$, and $r_{\mathrm{OTS}}\simeq7.48~\Aa$. The variational flux minimum
is not a separate statistical surface; it is the dividing surface spanned by the tight periodic orbit, the
orbit that gates entry to the well. In one sentence: a localized interior maximum of the transverse
stiffness reshapes roaming, gating inner capture and displacing roaming outward, by placing a deep
entropic bottleneck at an interior radius. That bottleneck is the tight transition state, an
unstable periodic orbit whose dividing surface is the variational minimum-flux surface, barrierless along the reaction
coordinate. Its entropic character it shares with the Chesnavich tight transition state; what the
interior maximum supplies is the concentration of hindrance strength at the gating radius that
makes it control capture. The mechanism is the angular
analogue of Makarov's transverse-stiffening entropic barrier, here placed in a roaming Hamiltonian
with a long-range tail and an outer centrifugal transition state, and the roaming it admits is
nonstatistical.

\appendix

\section{Computational methods}
\label{app:methods}

\paragraph{Trajectory ensemble.}
Trajectories of Eq.~\eqref{eq:Hdesign} are integrated with a fixed-step fourth-order Runge--Kutta (RK4) scheme
($\Delta t=0.005$ in model time units, $\tau=0.0489$~ps per unit) at $E=2$~kcal/mol. The
fixed-energy ensemble is a $240\times240$ grid in $(\theta_0,p_{\theta,0})$ launched inward from
$r_0=6~\Aa$. At each grid point the inward radial momentum is fixed by the energy constraint,
\begin{equation*}
p_r=-\sqrt{2\mu\bigl[E-V_{\mathrm{rad}}(r_0)-\tfrac12 B(r_0)(1-\cos2\theta_0)-\tfrac12
a(r_0)p_{\theta,0}^2\bigr]},
\end{equation*}
and grid points for which the radicand is non-positive (energetically
inaccessible) are removed; reported fractions are over the energetically accessible set. A trajectory
is integrated until it escapes, crossing $r=8~\Aa$ outward, or reaches the time limit
$t_{\max}=320$ model units ($\approx15.6$~ps).

\paragraph{The meaning of ``reactive.''}
Trajectories are not stopped at the inner boundary. Each is integrated until it escapes (crosses
$r=8~\Aa$ outward) or reaches the time limit $t_{\max}=320$ model units, and its closest approach
$r_{\min}$ and its crossings of $r=3.5~\Aa$ are recorded along the way. A trajectory is
\emph{reactive} if $r_{\min}<1.6~\Aa$, that is, if it enters the well, and \emph{roaming} if it
crosses $r=3.5~\Aa$ three or more times in either direction. Because capture is registered by the
closest approach, the inner-capture fractions of Sec.~\ref{sec:matched} do not depend on whether
integration is continued past the well: stopping there would affect only trajectories that enter
and later re-emerge. Continuing resolves that small class ($0.13\%$ of trajectories enter the well,
cross back outward, and roam, the roaming-reactive class, alongside the $6.04\%$ that roam without
entering) and both come from the same run. ``Reactive'' therefore denotes entry into the well, not
committed product formation: the flow is volume-preserving and a captured trajectory may in
principle redissociate.

\paragraph{Switching-parameter dependence.}
The Chesnavich lock width $\alpha$ controls how early the angular hindrance switches off. The
comparisons of Sec.~\ref{sec:matched} use the standard value $\alpha=1$; the early-switching case
$\alpha=4$ is reported there to show that the roaming fraction is parameter-dependent while
the inner-well gating is not; the outward redistribution is demonstrated explicitly only for
the $\alpha=1$ classifier-radius sweep.

\paragraph{Exact directional flux.}
The flux Eq.~\eqref{eq:flux} and confinement Eq.~\eqref{eq:Cfac} are evaluated by adaptive
quadrature in $\theta$ on a dense grid in $R$; the integrand's positive-part support is resolved by
refining near the turning angles. The reported minima are stable to grid refinement at the
$\sim0.01~\Aa$ level, which is the resolution to which $\Rstar$ is quoted.

\paragraph{Outer boundary.}
Capture and classifier statistics use an outer removal radius $r=8~\Aa$; the orbiting transition
state and the effective-potential construction of Sec.~\ref{sec:dynamical} were checked against an
extended boundary $r=12~\Aa$, with no change to $r_{\mathrm{OTS}}$ or to the reported fractions
beyond the last quoted digit.

\paragraph{Numerical validation.}
The reported fractions are converged in grid, step, and integration time. On the Chesnavich
channel the inner-capture fractions move from $57.11\%/14.68\%$ (lock/ridge) on a $120\times120$
launch grid to $57.27\%/14.75\%$ on the $240\times240$ production grid, a shift below $0.1$
percentage points; the capture fraction is identical to two decimals at $\Delta t=0.01$, $0.005$,
and $0.0025$. The fixed-step RK4 integrator conserves energy to $|H-E|\lesssim2\times10^{-7}$
kcal/mol over complete trajectories, and at $t_{\max}=320$ model units ($\approx15.6$~ps) only
$0.085\%$ of trajectories ($49$ of $57{,}600$) remain unclassified inside the interaction region.

\paragraph{Tight-orbit dividing-surface flux.}
The tight periodic orbit is located by a Newton search on the $\sin\theta=0$ surface of section
(converged initial condition $(r,p_r,\theta,p_\theta)=(2.4689,0,0,4.3505)$, period $T=0.5134$ at
$E=2$). The flux through the dividing surface it spans equals its action $\oint p\,dq$,
accumulated along one period by RK4 at $\Delta t=T/1.2\times10^{5}$; the orbit closes to
$|y(T)-y(0)|<10^{-4}$ with $|H-E|\le10^{-4}$ along it. The resulting action, $2.902$, agrees with
the per-window fixed-radius flux minimum $\Phi_{R^*}/2=2.907$ to $0.15\%$
(Fig.~\ref{fig:orbits}c).

\section{Gap-time analysis}
\label{app:gap}

The \emph{gap time} of a trajectory is the residence time between one inward crossing of the launch
surface $r_0=6~\Aa$ and the next outward crossing. Its distribution is a classical diagnostic of
statistical versus nonstatistical decay: in the statistical (RRKM) picture escape is memoryless, and
the survival function $S(t)$, the fraction of the entering ensemble still inside at time $t$, is a
single exponential set by one rate-limiting bottleneck. A departure from a single exponential
indicates that the escape is not memoryless: the residence times carry structure that a single rate
does not capture, either a tail heavier than exponential (trapping and recrossing) or a short-time
shoulder with a lighter tail (fast, direct transits on a characteristic crossing time). The gap-time
formulation is due to Thiele~\cite{thiele1962}, following Slater's dynamical theory of
unimolecular reactions~\cite{slater1956,slater1959}, and was cast in the phase-space
transition-state setting by Ezra, Waalkens and Wiggins~\cite{ezra2009gap}; Maugui\`ere, Collins, Ezra, Farantos and
Wiggins applied it to the roaming region of the Chesnavich model and likewise found nonstatistical
behavior~\cite{mauguiere2014jcp}.

The criterion deserves emphasis. Statisticality is a hypothesis about dynamics---memoryless escape
at a single rate---and the gap-time distribution tests it directly on the trajectories themselves.
Nonstatisticality in roaming is sometimes framed instead in terms of ergodic exploration of the
long-range region~\cite{suits2020}; but the energy shell here is unbounded and trajectories escape
through the radical channel, so ergodicity in its standard sense is not defined for this flow,
whereas the gap-time test is constructed for precisely such open systems.

We compute $S(t)$ on the same fixed-energy ensemble used throughout, the $240\times240$ inward
grid launched from $r_0=6~\Aa$ (Appendix~\ref{app:methods}), so that the diagnostic describes the
very trajectories whose capture and roaming Secs.~\ref{sec:matched} and~\ref{sec:roaming}
characterize, not a separately launched set. For each nonreactive trajectory we record the time
from its inward launch at $r_0=6~\Aa$ until it escapes, crossing $r=8~\Aa$ outward. Because
$r_0=6~\Aa$ lies inside the orbiting transition state ($r_{\mathrm{OTS}}\simeq7.48~\Aa$), these are
conditional-on-entry residence times, not full collision lifetimes; the non-exponential character
found below holds for other launch surfaces as well, so $r_0=6~\Aa$ is fixed by consistency with
the rest of the paper rather than to obtain the result. Two caveats on the construction itself.
Conditioning on the nonreactive outcome means this is not the gap-time distribution of the
complete incoming flux through a dividing surface; it is the conditional distribution of
nonreactive return times, and it is that distribution we test. And part of the short-time
behavior is deterministic: no trajectory can exit before the minimum transit time from $r_0$ in
and back out, so the flat shoulder at small $t$ is guaranteed, and the informative comparison
with an exponential is the shape of the decay beyond that direct-transit interval.

\begin{figure}[t]
\centering
\includegraphics[width=0.62\textwidth]{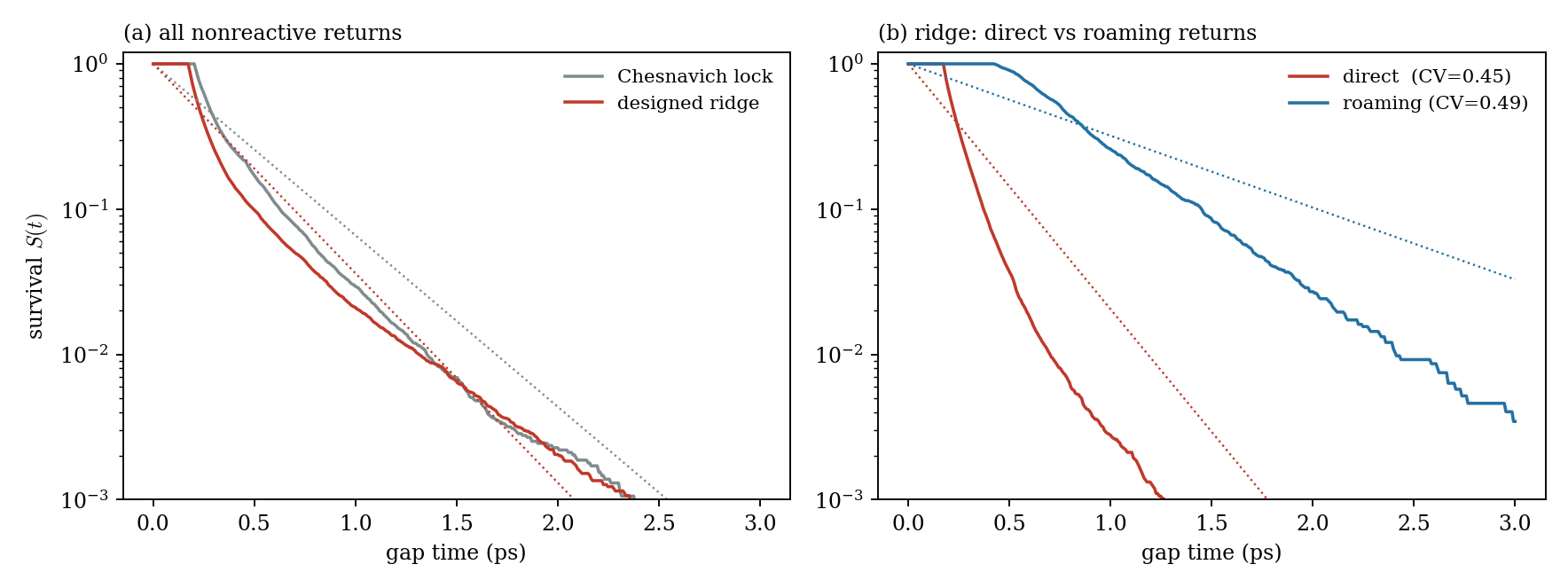}
\caption{Nonreactive gap-time survival ($r_0=6~\Aa$ launch, conditional on inward entry).
(a)~Chesnavich lock and entropic ridge; dotted curves are matched single exponentials. Both
distributions are narrower than exponential, with a delayed onset set by the minimum transit
time. (b)~The ridge ensemble resolved into direct and roaming-classified returns: the two
populations are individually narrow (coefficient of variation CV $=0.45$ and $0.49$) and well separated in time; the full
distribution is their bimodal mixture, inconsistent with memoryless single-rate decay.}
\label{fig:gap}
\end{figure}

Figure~\ref{fig:gap}a shows $S(t)$ for the Chesnavich lock and the entropic ridge, each with the
matched single exponential overlaid. Both are manifestly non-exponential and \emph{narrower} than
exponential: coefficient of variation (CV, the ratio of standard deviation to mean) $0.68$ for the lock and $0.75$ for the ridge, against $1$
for a memoryless decay, with tails lighter than exponential beyond the transit shoulder.
Figure~\ref{fig:gap}b resolves the ridge ensemble into its direct and roaming-classified
components, and this is the sharper statement. The two populations are well separated---mean
return times $0.257$~ps (direct) and $0.878$~ps (roaming)---and each is itself narrow (CV $0.45$
and $0.49$); the lock decomposes the same way ($0.302$ and $0.775$~ps, CV $0.44$ and $0.49$). The
full distribution is therefore a bimodal mixture of fast direct returns and slower roaming
returns, not a memoryless decay at any single rate. In particular the roaming-classified
subset---the population whose statisticality is at issue---has return times tightly clustered
about a characteristic value where a statistical (RRKM) escape would give CV $=1$ and an
exponential tail: the roaming is nonstatistical in the direct sense that its residence times are
inconsistent with memoryless single-rate escape. The entropic bottleneck sets how much enters,
not the pace of what crosses: the transit-time-versus-rate decoupling Makarov identifies for an
entropic barrier. The mean nonreactive gap times over the full ensembles are $0.368$~ps (lock)
and $0.301$~ps (ridge).

\section*{Data availability}
This study reports no experimental data. All results were generated by trajectory
integration of the classical Hamiltonian models defined in
Sections~\ref{sec:chesnavich} and \ref{sec:design}, with the parameters of
Table~\ref{tab:param} and the integrator of Appendix~\ref{app:methods}. The
scripts that reproduce every figure and reported quantity are available from the author upon reasonable request.


\end{document}